# Smooth Entropy Transfer of Quantum Gravity Information Processing

Laszlo Gyongyosi

[1] Quantum Technologies Laboratory, Department of Telecommunications
*Budapest University of Technology and Economics*
2 Magyar tudosok krt, Budapest, *H*-1117, Hungary
[2] Information Systems Research Group, Mathematics and Natural Sciences
*Hungarian Academy of Sciences*
Budapest, *H*-1518, Hungary

gyongyosi@hit.bme.hu

**Abstract**

We introduce the term smooth entanglement entropy transfer, a phenomenon that is a consequence of the causality-cancellation property of the quantum gravity environment. The causality-cancellation of the quantum gravity space removes the causal dependencies of the local systems. We study the physical effects of the causality-cancellation and show that it stimulates entropy transfer between the quantum gravity environment and the independent local systems of the quantum gravity space. The entropy transfer reduces the entropies of the contributing local systems and increases the entropy of the quantum gravity environment. We discuss the space-time geometry structure of the quantum gravity environment and the local quantum systems. We propose the space-time geometry model of the smooth entropy transfer. We reveal on a smooth Cauchy slice that the space-time geometry of the quantum gravity environment dynamically adapts to the vanishing causality. We define the corresponding Hamiltonians and the causal development of the quantum gravity environment in a non-fixed causality structure. We prove that the Cauchy area expansion, along with the dilation of the Rindler horizon area of the quantum gravity environment, is a strict corollary of the causality-cancellation of the quantum gravity environment.

# 1 Introduction

The theory of quantum gravity [1-10], [12-32], [38-44], [46-58] fuses the non-fixed causality structure of general relativity and the quantum uncertainty of quantum mechanics into a unified framework [1-9]. In a quantum gravity environment, processes and events are causally non-separable because the term *time* and the time steps have no interpretable meaning in a non-fixed causality structure. This space-time structure allows us to perform quantum gravity computations and to build quantum gravity computers that are not just equipped with the power of quantum computations but are also operating on a non-fixed causality structure [4].

In this work, we are focusing on the information processing capabilities of the quantum gravity space, precisely on the physical fundaments of *quantum gravity information processing*. The vanishing causality of the quantum gravity space injects additional resources into the information processing structure. In particular, physically separated, local information carriers (referred as *subsystems* throughout) in the quantum gravity space can transfer information among one another through the quantum gravity space, which has been recently called the *information resource pool property* of the quantum gravity space [9]. The information resource pool property allows for a remote Alice (sender) and Bob (receiver) to simulate their remote outputs in a probabilistic way, through their local environment state. In particular, the reason behind this phenomenon is that the quantum gravity space, in fact, acts as an additional resource for the parties; in particular, it is operating as an information transfer device. We reveal that the vanishing causality structure of the quantum gravity space, which we refer by the *causality-cancellation* of the quantum gravity environment, has physical consequences on the quantum gravity environment and on the subsystems. Specifically, the causality-cancellation removes the causal dependencies of the contributing subsystems within the quantum gravity space and makes the causality structure into a non-fixed. This phenomenon is acclaimed here as an essential ingredient of quantum gravity information processing. Our main aim was to reveal those physical attributes that characterize the information processing structure of the quantum gravity in a non-fixed causality as well as to provide a plausible model for it. In particular, the causality-cancellation stimulates the phenomenon of *smooth entanglement entropy transfer*, which increases the entanglement entropy of the quantum gravity environment and decreases the entanglement entropies of the local subsystems that are communicating with one another through the quantum gravity space.

We reveal the physical characteristic of the smooth entropy transfer of the quantum gravity space and study the changes of the geometry of the space-time structure. We prove that the causality-cancellation is precisely rooted in information transfer between the quantum gravity environment and the contributing subsystems within the quantum gravity space. Because the quantum gravity space itself has quantum informational origin, it implies that the entropic gravity is related to the entanglement entropy [11], measure of which is also essential in our derivations. It also has been found that the Einstein equations are related to information transfer and also connected to the laws of thermodynamics [11]. Our results use these perceptions and also have further consequences in the modeling of the entropy transfer of quantum gravity information processing. To reveal the characteristics of the vanishing causality structure of quantum gravity, we exploit and fuse the results of continuum quantum field theory (QFT) [12-32], [38-44], [46-58] with quantum Shannon theory [33-37], [45-46].

The space-time geometry of the quantum gravity space is analyzed through a smooth Cauchy slice of some smooth background geometry. It allows us to study the changes of the space-time geometry in a vanishing causality structure through the entanglement entropy function. In particular, this connection arises from the fact that a Cauchy region of the quantum gravity environment on a Cauchy slice is derived from the entanglement entropy and also leads to a well-tractable area law [53] formula. Our results are further verified by the fact that it also has been demonstrated that the entanglement entropy function can be proposed as a probe of the space-time architecture of the quantum gravity space [10]. We show that the causality-cancellation restructures the Cauchy areas in a smooth space-time; in particular, the entropy transfer increases the Rindler horizon[*] area seen by an accelerating Rindler observer (referred to as the *Rindler horizon area of the quantum gravity environment* throughout) in a regularized region on a smooth Cauchy slice.

This paper is organized as follows. Section 2 summarizes some preliminary findings. Section 3 proposes the geometry structure of quantum gravity information processing. Section 4 discusses the causality-cancellation of the quantum gravity environment and the phenomenon of the smooth entropy transfer. Section 5 concludes the results. Supplementary information is included in the Appendix.

## 2 Preliminaries

### 2.1 Basic Terms and Definitions

Let $|\psi\rangle$ refer to a pure, not degenerate ground state with density matrix $\rho = |\psi\rangle\langle\psi|$. For practical purposes, the density $\rho$ can be generalized into a thermal density matrix, as follows:

$$\rho = e^{\frac{-H}{T}}, \qquad (1)$$

where $H$ is the total Hamiltonian, and $T$ is the temperature.
The physical quantity is expressed as

$$\langle O \rangle = Tr(O \cdot \rho), \qquad (2)$$

where $Tr(\rho) = 1$ [60-63].

For a generic quantum system at a finite temperature $T_g$, the $\sigma_g$ density matrix of the system is mixed, thus for the canonical ensemble

$$\sigma_g = \tfrac{1}{Z} e^{-\eta H}, \qquad (3)$$

where $Z$ is the partition function, evaluated as

$$Z = Tr(e^{-\eta H}). \qquad (4)$$

---

[*] For an online tutorial on the Rindler horizon, see [59].

Bipartitioning $\rho$ into subsystems *A* and *B*, the entanglement entropy $S_A$ of subsystem *A* can be evaluated by the reduced density matrix, as:

$$\rho_A = Tr_B(\rho), \tag{5}$$

which is given by tracing out the degrees of freedom in subsystem *B*.

For a pure state $\rho$ with Hilbert space $\mathcal{H}_{tot} = \mathcal{H}_A \otimes \mathcal{H}_B$, the entanglement entropies of the partitions *A* and *B* of $\rho$,

$$S_A = S_B, \tag{6}$$

which follows from the Schmidt decomposition of the pure $\rho$.

The reduced density matrix $\rho_A$ is both Hermitian and semi-definite,

$$\rho_A = e^{-H_A}, \tag{7}$$

where $H_A$ is the modular Hamiltonian.

From the reduced density $\rho_A$, the entanglement entropy is expressed by the von Neumann entropy as

$$S_A = -Tr(\rho_A \log_2 \rho_A). \tag{8}$$

In particular, in continuum QFT, assuming a *d*-dimensional space-time structure, the entanglement entropy of $\rho_A$ can be defined on a smooth $\mathcal{C}$ Cauchy slice by partitioning the space at time-slice $t = 0$ into region *A* and its complement region $\bar{A}$. These are spatial regions on a smooth Cauchy slice $\mathcal{C}$ of some smooth background geometry $\mathcal{B}$ [10], [13-14].

Specifically, the spatial regions *A* and $\bar{A}$ are separated by an entangling surface $\Sigma_A$, which is defined as follows:

$$\Sigma_A = \partial A = \partial \bar{A}. \tag{9}$$

For a region *A*, the reduced density $\rho_A$ can be determined by integrating out the degrees of freedom in the complementary region $\bar{A}$, then the von Neumann entropy $S_A$ of $\rho_A$ can be calculated using Equation (8). However, because the von Neumann entropy function is divergent in continuum theories, it requires a short-distance cutoff regulator $\delta$. In particular, the length scale $\delta$ acts as an UV (ultraviolet) regulator [10], [13].

In particular, at a short-distance cutoff regulation $\delta$ of the field theory, the entanglement entropy of region *A* on a smooth Cauchy slice $\mathcal{C}$ in the *d*-dimensional space-time is evaluated (*geometric entropy*) as follows:

$$S_A = \frac{C}{\delta^{d-1}} A + \chi, \tag{10}$$

where *C* is a dimensionless constant determined by the actual continuum field theory, $\delta^{d-1}$ is a UV cutoff (lattice spacing), and $\chi$ refers to the subleading terms [13].

Specifically, the entanglement entropy satisfies the strong sub-additivity; thus, for subsystems $A_1$ and $A_2$,

$$S_{A_1} + S_{A_2} \geq S_{A_1 \cap A_2} + S_{A_1 \cup A_2}. \tag{11}$$

Considering an arbitrary QFT in the *d*-dimensional Minkowski space $R^{1,d-1}$, the entanglement entropy can be calculated on a $\mathcal{C}$ Cauchy slice, across a spherical surface $S^{d-2}$.

The $D_A$ causal development of $A$ is the set of all $p_i$ points, for which all causal curves through $p_i$ intersects the region $A$ on the smooth Cauchy slice $\mathcal{C}$ [10], expressed as:

$$D_A = \bigcup_{\forall p_i} \{A \cap p_i\}. \tag{12}$$

The causal development $D_A$ of a region $A$ inside the spherical surface $S^{d-2}$ can be mapped onto a space $\mathcal{S} = R \times H^{d-1}$ [14]. In particular, the causal development of a Minkowski vacuum state $\rho_0$ is the Rindler wedge $L$ [10].

In a quantum gravity setting, for a sufficiently large geometric scale $d_{geom}$ in a smooth background space-time, that is, for sufficiently large Cauchy region $A$ on $\mathcal{C}$, the geometric entanglement entropy between the degrees of freedom can be expressed as follows:

$$S_A^{(d_{geom})} = \frac{2\pi}{l^{d-2}} A^{(d_{geom})} + \dots, \tag{13}$$

where $l^{d-2}$ is the Planck scale in $d$ space-time dimensions [10], evaluated as:

$$l^{d-2} = \frac{8\pi G \hbar}{c^3}, \tag{14}$$

where $G$ is Newton's constant, $\hbar$ is the reduced Planck constant, and $c$ is the speed of light in a vacuum.

If the scale on $\mathcal{C}$ is regulated to a short-distance cutoff $\delta \ll d_{geom}$, then the entanglement entropy can be evaluated as follows:

$$S_A^{(\delta)} = c_0 \frac{\tau^{d-2}}{\delta^{d-2}} + c_2 \frac{\tau^{d-4}}{\delta^{d-4}} + \dots, \tag{15}$$

where $\tau$ is some macroscopic scale that characterizes the entangling surface $\Sigma_A$, and $c_k$ is the dimensionless coefficient (sensitive to the regulator of the field theory), which after some calculations results in the area law formula for a Cauchy region $A$ as

$$S_A = \frac{\tilde{c}_0}{\delta^{d-2}} A, \tag{16}$$

where $\tilde{c}_0$ is also a regulator-dependent coefficient [10], [14], [53].

At a regulation $d_A \ll d_{geom}$, a space-time region $\Gamma_A$ of size $d_A$ near the entangling surface $\Sigma_A$ of Cauchy region $A$ can be defined, such that

$$\delta \ll d_A \ll d_{geom}. \tag{17}$$

In particular, in a space-time region $\Gamma_A$, the space-time looks like a flat space [10], and the $D_A^{(d_A)}$ causal development of $A$ at $d_A$ coincidences to a Rindler horizon $L_A$; thus,

$$D_A^{(d_A)} = \partial D_A = L_A, \tag{18}$$

where $D_A$ is the causal development of $A$ at $d_{geom}$ [10].

Note, throughout the manuscript, the area of the entangling surface $\Sigma_A$ will be referred via the Cauchy area $A$, and the quantities of $D$ and $L$ will be used as spatial area terms.

## 2.2 Quantum Gravity Information Processing

In this section, we briefly summarize the information processing model of the quantum gravity environment from [9].

Let $\mathcal{M}_A$ and $\mathcal{M}_B$ be two independent, physically separated, local CPTP (completely positive trace preserving) maps within the quantum gravity environment $\mathcal{G}_E$ (the results can be generalized for an arbitrary number of $\mathcal{M}$ within $\mathcal{G}_E$). The maps are initialized with uncorrelated inputs $A_1$ and $A_2$. The local input is denoted by $A_i$, and the local outputs and environments are denoted by $B_i$ and $E_i$, respectively. The remote output is referred to as $B_j$, $j \neq i$. The inputs can convey classical or quantum information, both the same type. A local $\mathcal{M}_i$ can be decomposed into the local logical channel $\mathcal{N}_{A_i B_i}$, which exists between the input $A_i$ and the output $B_i$, and the local complementary channel $\mathcal{N}_{A_i E_i}$, which connects the input $A_i$ with the local environment state $E_i$. Both $\mathcal{N}_{A_i B_i}$ and $\mathcal{N}_{A_i E_i}$ are assumed to be qubit maps.

The local environment state and remote outputs $E_1$ and $B_2$ of $\mathcal{M}_A$ and $\mathcal{M}_B$ are entangled with the quantum gravity environment state $\mathcal{G}_E$, formulating an entangled density matrix $\rho_{\mathcal{G}_E E_1 B_2}$, in which $E_1$ is separable from $\mathcal{G}_E B_2$, $B_2$ is separable from $\mathcal{G}_E E_1$, and $\mathcal{G}_E$ is entangled with $E_1 B_2$. Without loss of generality, systems $\rho_{\mathcal{G}_E E_1 B_2}$ and $\rho_{\mathcal{G}_E E_1 B_2}$ formulate the following density matrix:

$$\rho = \tfrac{1}{2}\rho_{\mathcal{G}_E E_1 B_2} + \tfrac{1}{2}\rho_{\mathcal{G}_E E_2 B_1}. \tag{19}$$

In particular, focusing on the tripartite system $\rho_{\mathcal{G}_E E_1 B_2}$, the following conditions have to be satisfied for the partitions $\mathcal{G}_E - E_1 B_2$, $E_1 - \mathcal{G}_E B_2$, and $B_2 - \mathcal{G}_E E_1$. Because the local subsystems $E_1$ and $B_2$ have to be separable from the partitions $\mathcal{G}_E B_2$ and $\mathcal{G}_E E_1$, in this tripartite system, only the quantum gravity environment $\mathcal{G}_E$ can be entangled with $E_1 B_2$, and all other partitions have to be separable with respect to $E_1$ and $B_2$. From these, it clearly follows that the partitions $E_1 - \mathcal{G}_E B_2$ and $B_2 - \mathcal{G}_E E_1$ have to be separable, and $\mathcal{G}_E - E_1 B_2$ has to be entangled. For further details and derivation, see [9].

The entangled structure of quantum gravity environment $\mathcal{G}_E$ is depicted in Fig. 1 [9]. The information transmission is realized through the partition $\mathcal{G}_E - E_i B_j$, $i \neq j$.

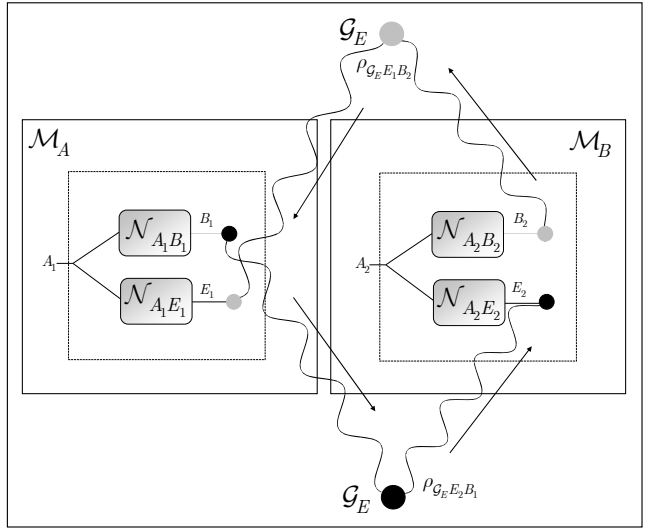

**Figure 1.** The density matrix $\rho = \frac{1}{2}\rho_{\mathcal{G}_E E_1 B_2} + \frac{1}{2}\rho_{\mathcal{G}_E E_2 B_1}$. The local environment state $E_i$ and the remote output $B_j$ of $\mathcal{M}_A$ and $\mathcal{M}_B$ are entangled with the quantum gravity environment state $\mathcal{G}_E$, via the partition $\mathcal{G}_E - E_i B_j$, $j \neq i$. The entanglement between the local environments and the quantum gravity environment allows the parties to simulate locally the remote outputs from their local environment. (The wavy lines illustrate the entanglement; the arrow refers to the direction of the information flow.)

As it will be proposed in Section 4, the non-fixed causality between density matrices $\rho_{\mathcal{G}_E E_1 B_2}$ and $\rho_{\mathcal{G}_E E_2 B_1}$ arises from the causality-cancellation of $\mathcal{G}_E$. Specifically, the phenomenon of causality-cancellation is based on the information resource pool property of the quantum gravity environment. The information resource pool property [9] of the quantum gravity space allows the parties to perform the so-called remote simulation in a vanishing causality structure. In particular, in a plausible manner, the outputs $B_{1,2}$ of the local maps $\mathcal{M}_A$ and $\mathcal{M}_B$ (at Alice and Bob) "write in" information into the resource pool medium, the quantum gravity environment $\mathcal{G}_E$. The information is "read out" from the pool $\mathcal{G}_E$ by the local environments $E_{2,1}$. The quantum gravity environment, in fact, acts as a shared multiple-access medium between Alice and Bob, with parallel information reading and writing capabilities. For an exact derivation of a density matrix $\rho_{\mathcal{G}_E E_i B_j}$, see [9].

In Fig. 2, the causal directions of the information flow from Alice to Bob, and from Bob and Alice are depicted by $A \rightarrow B$ and $B \rightarrow A$ on the red dashed arrows, respectively. The subsystems $B_1 - E_2$ and $B_2 - E_1$ are causally connected because in a fixed causality, $E_2$ results from $B_1$, and $E_1$ results from $B_2$. In a vanishing causality structure, both directions, depicted by $A \rightleftarrows B$, are realized in parallel through the quantum gravity environment $\mathcal{G}_E$ without causality relations between $B_{1,2}$ and $E_{2,1}$, via the density matrix of Equation (19).

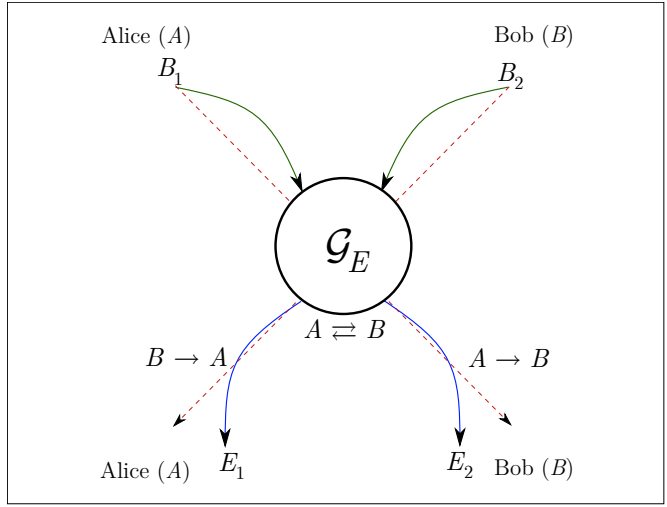

**Figure 2.** The causality-cancellation of the quantum gravity environment. The causality-cancellation results in the parallel realization, $A \rightleftarrows B$, of causal directions $A \rightarrow B$ and $B \rightarrow A$ ($A$, Alice; $B$, Bob).

As it will be proposed in Section 4, the causality-cancellation of the quantum gravity space $\mathcal{G}_E$ has physical consequences because it increases the entropy of the quantum gravity environment and reduces the entropies of the subsystems $E_i$ and $B_j$.

## 3  Geometry of Quantum Gravity Information Processing

**Theorem 1** (Information processing structure of the quantum gravity environment on a smooth Cauchy slice.) *Considering a region $\mathcal{G}_E$ for the quantum gravity environment on a smooth Cauchy slice $\mathcal{C}$, the local systems $E_i$ and $B_j$ lie in the $\overline{\mathcal{G}}_E$ complementary region of $\mathcal{G}_E$. The entangling surface $\Sigma_{\mathcal{G}_E}$ entangles region $\mathcal{G}_E$ with the complementary region $\overline{\mathcal{G}}_E = E_i B_j$, generating the entangled partition $\mathcal{G}_E - E_i B_j$ of the density matrix $\rho_{\mathcal{G}_E E_i B_j}$. The reduced density matrix $\rho_{\mathcal{G}_E}$ defines $D_{\mathcal{G}_E}$. At a regulator $d_{\mathcal{G}_E}$, $\delta \ll d_{\mathcal{G}_E} \ll d_{geom}$, $D_{\mathcal{G}_E}^{(d_{\mathcal{G}_E})}$ is a Rindler horizon $L_{\mathcal{G}_E}$, where $d_{geom}$ is a sufficiently large geometric scale on $\mathcal{C}$.*

*Proof.*
The proofs throughout assume the use of two independent $\mathcal{M}_A$ and $\mathcal{M}_B$ maps within the quantum gravity space, with the density matrix (19). The proof uses the preliminaries and notations of Section 2.2. The results of the proofs can be extended and generalized for an arbitrary setting. Let $\mathcal{C}$ be a smooth Cauchy slice of a smooth background geometry $\mathcal{B}$. The information processing structure consists of the $\mathcal{G}_E$ quantum gravity environment and two independent local quantum systems $E_i$ and $B_j$ within the quantum gravity environment. The system $\mathcal{G}_E$ is entangled with $E_i$ and $B_j$ through the partition $\mathcal{G}_E - E_i B_j$, whereas the other partitions of $\rho_{\mathcal{G}_E E_i B_j}$ (see Equation (19)) are unentangled (see Section 2 and [9]).

The $\mathcal{G}_E$ quantum gravity environment has an area $\mathcal{G}_E$ on a $\mathcal{C}$ Cauchy slice. The boundary of the area of $\mathcal{G}_E$ defines an entangling surface $\Sigma_{\mathcal{G}_E}$ as follows:

$$\begin{aligned} \Sigma_{\mathcal{G}_E} &= \partial \mathcal{G}_E \\ &= \partial \bar{\mathcal{G}}_E. \end{aligned} \qquad (20)$$

The area of $\Sigma_{\mathcal{G}_E}$ will be referred by the Cauchy area $\mathcal{G}_E$ throughout. Specifically, the entanglement of $\mathcal{G}_E - E_i B_j$ is modeled on $\mathcal{C}$, as follows. The $\Sigma_{\mathcal{G}_E}$ entangling surface of $\mathcal{G}_E$ entangles the $\mathcal{G}_E$ gravity environment region with the regions of the systems $E_i$ and $B_j$ on $\mathcal{C}$, which lie in the complementary area of $\mathcal{G}_E$ on $\mathcal{C}$. The complementary region of $\mathcal{G}_E$ on $\mathcal{C}$ is depicted by $\bar{\mathcal{G}}_E$, as:

$$\bar{\mathcal{G}}_E = E_i B_j, \qquad (21)$$

because only partition $\mathcal{G}_E - E_i B_j$ contains entanglement in the full density matrix of $\rho_{\mathcal{G}_E E_i B_j}$.

Let $\mathcal{H}_{tot}$ the total Hilbert space of the density matrix $\rho_{\mathcal{G}_E E_i B_j}$. Partitioning the density matrix $\rho_{\mathcal{G}_E E_i B_j}$ into two parts, $\mathcal{G}_E$ and $\bar{\mathcal{G}}_E = E_i B_j$, the density $\rho_{\mathcal{G}_E E_i B_j}$ can be rewritten as $\rho_{\mathcal{G}_E \bar{\mathcal{G}}_E}$, and the $\mathcal{H}_{tot}$ total Hilbert space becomes factorized:

$$\mathcal{H}_{tot} = \mathcal{H}_{\mathcal{G}_E} \otimes \mathcal{H}_{\bar{\mathcal{G}}_E}. \qquad (22)$$

The partition $\mathcal{G}_E - E_i B_j$ is modeled via Cauchy regions $\mathcal{G}_E$ and $\bar{\mathcal{G}}_E$ on the smooth Cauchy slice $\mathcal{C}$, and the $S_{\mathcal{G}_E}$, and $S_{\bar{\mathcal{G}}_E}$ entanglement entropy of $\mathcal{G}_E$ and $\bar{\mathcal{G}}_E$ are evaluated via the (Hermitian and semi-definite) reduced density matrices:

$$\rho_{\mathcal{G}_E} = Tr_{\bar{\mathcal{G}}_E}\left(\rho_{\mathcal{G}_E \bar{\mathcal{G}}_E}\right), \qquad (23)$$

and

$$\rho_{\bar{\mathcal{G}}_E} = Tr_{\mathcal{G}_E}\left(\rho_{\mathcal{G}_E \bar{\mathcal{G}}_E}\right). \qquad (24)$$

Thus, $S_{\bar{\mathcal{G}}_E}$ of $\rho_{\bar{\mathcal{G}}_E}$ can be rewritten as:

$$S_{\bar{\mathcal{G}}_E} = S_{E_i} + S_{B_j}. \qquad (25)$$

where $S_{E_i} = -Tr\left(\rho_{E_i} \log_2 \rho_{E_i}\right)$ and $S_{B_j} = -Tr\left(\rho_{B_j} \log_2 \rho_{B_j}\right)$.

The density matrix $\rho_{\mathcal{G}_E}$ of $\mathcal{G}_E$ defines the $D_{\mathcal{G}_E}$ casual development on $\mathcal{C}$, as follows. Let $H_{\mathcal{G}_E}$ be the Hermitian operator (also referred to as *entanglement Hamiltonian*) of $\mathcal{G}_E$, then the reduced density matrix $\rho_{\mathcal{G}_E}$ of $\mathcal{G}_E$ is expressed as:

$$\rho_{\mathcal{G}_E} = e^{-H_{\mathcal{G}_E}}, \qquad (26)$$

which controls the system through the $D_{\mathcal{G}_E}$ casual domain. $\rho_{\mathcal{G}_E}$ is a reduced density matrix that results from the remaining degrees of freedom in Cauchy area $\mathcal{G}_E$, after the degrees of freedom have been integrated out from the Cauchy region $\bar{\mathcal{G}}_E$.

The causal development of the reduced density matrix $\rho_{\mathcal{G}_E}$ is as follows:
$$D_{\mathcal{G}_E} = \bigcup_{\forall p_i} \{\mathcal{G}_E \cap p_i\}. \tag{27}$$

In particular, at a sufficiently large geometric length scale $d_{geom} \gg \delta$ for Cauchy area $\mathcal{G}_E$, where $\delta$ is the short-distance cutoff regulator, and the $H_{\mathcal{G}_E}^{(d_{geom})}$ Hamiltonian is a non-local operator [10], expressed precisely as:
$$\begin{aligned} H_{\mathcal{G}_E}^{(d_{geom})} &= \int d^{d-1}x\, \gamma_1^{\mu\nu}(x) T_{\mu\nu} \\ &+ \int d^{d-1}x \int d^{d-1}y\, \gamma_2^{\mu\nu;\rho\sigma}(x,y) T_{\mu\nu} T_{\rho\sigma} + \ldots, \end{aligned} \tag{28}$$

where $d$ is the space-time dimension, and $\gamma_1^{\mu\nu}(\cdot)$ and $T_{\mu\nu}, T_{\rho\sigma}$ are the evolution parameters of the schematic expression of the Hamiltonian operator.

Without loss of generality, at a distance scale $d_{geom} \gg \delta$, the areas on $\mathcal{C}$ are precisely evaluated as follows:
$$\mathcal{G}_E^{(d_{geom})} = \frac{l^{d-2}}{2\pi} S_{\mathcal{G}_E}^{(d_{geom})} + \ldots, \tag{29}$$

where $l^{d-2}$ is the Planck scale in $d$ space-time dimensions. The entropy is then calculated as follows:
$$S_{\mathcal{G}_E}^{(d_{geom})} = \frac{2\pi}{l^{d-2}} \mathcal{G}_E^{(d_{geom})} + \ldots. \tag{30}$$

In a simplified form, the entropy of $\mathcal{G}_E$ is $S_{\mathcal{G}_E} = -Tr\left(\rho_{\mathcal{G}_E} \log_2 \rho_{\mathcal{G}_E}\right)$.

In particular, the causal development $D_{\mathcal{G}_E}$ is a Rindler horizon [10] at a regularization $d_{\mathcal{G}_E}$ (denoted by $D_{\mathcal{G}_E}^{(d_{\mathcal{G}_E})}$), where
$$\delta \ll d_{\mathcal{G}_E} \ll d_{geom}, \tag{31}$$

because within a space-time in a region $\Gamma$ of size $d_{\mathcal{G}_E}$ near the entangling surface $\Sigma_{\mathcal{G}_E}$, the space-time looks like a flat space with a Rindler horizon $L_{\mathcal{G}_E}$; thus,
$$D_{\mathcal{G}_E}^{(d_{\mathcal{G}_E})} = L_{\mathcal{G}_E} = \delta D_{\mathcal{G}_E}. \tag{32}$$

Specifically, for $d_{\mathcal{G}_E}$, the density $\rho_{\mathcal{G}_E}$ has the Rindler Hamiltonian [10]; in particular,
$$\begin{aligned} H_{\mathcal{G}_E}^{(d_{\mathcal{G}_E})} &= 2\pi K_{\mathcal{G}_E} + c_{\mathcal{G}_E} \\ &= -2\pi \int_{x>0} d^{d-2}y\, dx\, (x T_{00}) + c_{\mathcal{G}_E}, \end{aligned} \tag{33}$$

where $K_{\mathcal{G}_E}$ are boost generators in direction $x$, and $c_{\mathcal{G}_E}$ is a constant to reach the unit trace in $\rho_{\mathcal{G}_E}$. For a regularized $d_{\mathcal{G}_E}$, we can further exploit that the entanglement entropy function yields a constant divergence for the Rindler horizon $L_{\mathcal{G}_E}$, and integrating over $\mathcal{G}_E$, the leading singularity will result in the area law [10].

From these arguments, the area $\mathcal{G}_E$ on $\mathcal{C}$ can be precisely expressed as follows:

$$\mathcal{G}_E^{(d_{\mathcal{G}_E})} = \frac{\delta^{d-2}}{\tilde{c}_0} S_{\mathcal{G}_E}^{(d_{\mathcal{G}_E})}, \qquad (34)$$

where the constant $\tilde{c}_0$ depends on the actual regulator of the field theory; thus,

$$S_{\mathcal{G}_E}^{(d_{\mathcal{G}_E})} = \frac{\tilde{c}_0}{\delta^{d-2}} \mathcal{G}_E^{(d_{\mathcal{G}_E})}. \qquad (35)$$

The geometry structure on $\mathcal{C}$ at a non-vanishing causality between the local systems $E_i$ and $B_j$ at a geometric scale $d_{geom}$ is summarized in Fig. 3.

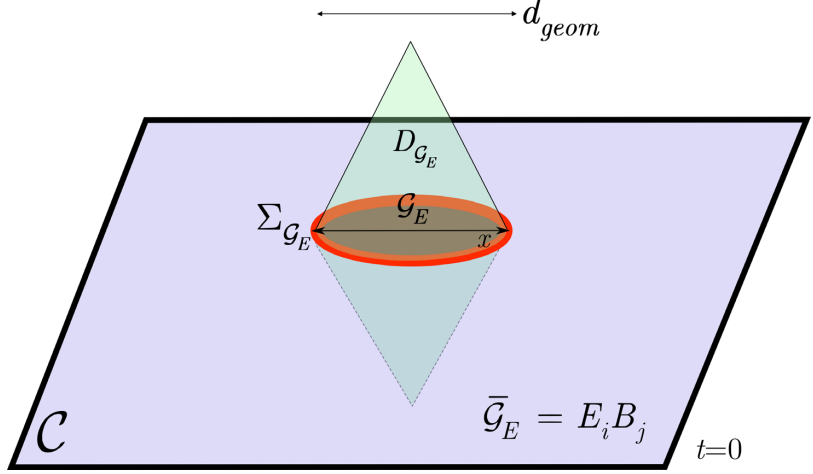

**Figure 3.** The entanglement between regions of the quantum gravity environment $\mathcal{G}_E$ and the local quantum systems $E_i$ and $B_j$ on a $\mathcal{C}$ Cauchy slice. The complementary region of $\mathcal{G}_E$ is $\bar{\mathcal{G}}_E = E_i B_j$. The Cauchy area $\mathcal{G}_E$ has an entangling surface $\Sigma_{\mathcal{G}_E} = \partial \mathcal{G}_E$ (red thick line), which entangles region $\mathcal{G}_E$ with region $\bar{\mathcal{G}}_E$, generating the entangled partition $\mathcal{G}_E - E_i B_j$ of the density matrix $\rho_{\mathcal{G}_E E_i B_j}$.

The smooth geometry on $\mathcal{C}$ defines the quantum gravity environment $\mathcal{G}_E$ with a non-zero entangling surface, see Equation (20), and the geometrical structure of $\rho_{\mathcal{G}_E E_i B_j}$ precisely generates entanglement only via the partition $\mathcal{G}_E - E_i B_j$ of $\rho_{\mathcal{G}_E E_i B_j}$. In particular, the structure of the complementary area $\bar{\mathcal{G}}_E$ allows information transfer through only the $\mathcal{G}_E - E_i B_j$ partition in the quantum gravity space, that is, $E_i - \mathcal{G}_E B_j = B_j - \mathcal{G}_E E_i = \varnothing$.

∎

As it will be revealed in Theorem 2, the geometry on $\mathcal{C}$ significantly changes if the causality between $E_i$ and $B_j$ is vanishing; however, the entanglement structure of $\rho_{\mathcal{G}_E E_i B_j}$ proven here remains untouched.

# 4 Smooth Entropy Transfer of Causality-Cancellation

**Proposition 1** (Causality-cancellation stimulates smooth entanglement entropy transfer between the quantum gravity environment and the local systems.) *The causality-cancellation of the quantum gravity environment $\mathcal{G}_E$ leads to the parallel realization $A \rightleftarrows B$ of the causal directions $A \to B$ and $B \to A$. The causality-cancellation of $\mathcal{G}_E$ stimulates smooth entropy transfer, which increases the entanglement entropy $S_{\mathcal{G}_E}$ of $\mathcal{G}_E$ and reduces the entropies of the local systems $E_i$ and $B_j$.*

The causality-cancellation of the quantum gravity space causes entropy transfer between the local systems and the quantum gravity space. In particular, this process is referred as the *entropy transfer* throughout. It restructures the geometry of $\rho_{\mathcal{G}_E E_i B_j}$ on the Cauchy slice $\mathcal{C}$. Assuming local maps $\mathcal{M}_A$ and $\mathcal{M}_B$ within the quantum gravity space, the causality-cancellation removes the causal dependencies between density matrices $\rho_{\mathcal{G}_E E_1 B_2}$ and $\rho_{\mathcal{G}_E E_2 B_1}$. On the other hand, the entanglement structure of the density matrix $\rho_{\mathcal{G}_E E_i B_j}$ shown in Section 2 remains untouched. These results are summarized in Theorem 2.

**Theorem 2** (Smooth entanglement entropy transfer of causality-cancellation.). *The entropy transfer increases the $S_{\mathcal{G}_E}$ to $S_{\mathcal{G}'_E} = S_{\mathcal{G}_E} + \left(S_{E_i^*} - S_0\right) + \left(S_{B_j^*} - S_0\right) = S_{\mathcal{G}_E} + S_{E_i^*} + S_{B_j^*} - 2S_0$ for all $i,j$ of $\rho$, where $i \neq j$, $S_{E_i^*} < S_{E_i}$, $S_{B_j^*} < S_{B_j}$, and $S_0 = -Tr\left(\rho_0 \log_2 \rho_0\right)$ is the entanglement entropy of a vacuum state $\rho_0$. The $S_{E'_i}$ and $S_{B'_j}$ operating entropies of $E'_i$ and $B'_j$ reduce to $S_{E'_i} = S_{E_i} - S_{E_i^*} + S_0$ and $S_{B'_j} = S_{B_j} - S_{B_j^*} + S_0$ in a non-fixed causality. The Cauchy area of $\mathcal{G}_E$ and the area of $D_{\mathcal{G}_E}^{\left(d_{\mathcal{G}_E}\right)}$ expand because of the smooth entropy transfer.*

*Proof.*
In the first part of the proof, we discuss the phenomenon of entropy transfer. In the second part, we study the effects of the non-fixed causality on the Cauchy slice. To make a clear distinction between the notes of Theorems 1 and 2, the elements of density matrix $\rho_{\mathcal{G}_E E_i B_j}$ at a non-fixed causality will be referred to as $\mathcal{G}'_E, E'_i, B'_j$. The proof assumes two local maps $\mathcal{M}_A$ and $\mathcal{M}_B$ within in the quantum gravity space; thus, $\rho$ is characterized by Equation (19). The results can be generalized for arbitrary number and setting of $\mathcal{M}$ within $\mathcal{G}_E$.

The entropies of $\mathcal{G}_E$, $E_i$, and $B_j$ in a fixed causality structure are to be $S_{\mathcal{G}_E}$, $S_{E_i}$, and $S_{B_j}$, respectively. In particular, for an inertial Rindler observer $\Theta_{in}$, the $\mathcal{G}_E$ gravity environment is a thermal system, i.e., a vacuum state $\rho_0$, with vacuum entropy $S_0 = -Tr\left(\rho_0 \log_2 \rho_0\right)$. While subsystems $E_i$ and $B_j$ are contributing with $\mathcal{G}_E$ through the process of causality-cancellation, the portions $E_i^*$ and $B_j^*$ are becoming a part of a thermal system of $\mathcal{G}_E$. The remaining "post causality-cancellation" systems are referred to as $\mathcal{G}_E'$, $E_i' = E_i - E_i^*$, and $B_j' = B_j - B_j^*$, respectively. Specifically, it introduces the entropic quantities of $S_{E_i^*} < S_{E_i}$, $S_{B_j^*} < S_{B_j}$, in which entropic portions vanish from $S_{E_i}$ and $S_{B_j}$ during the causality-cancellation.

In particular, the causality-cancellation between the quantum gravity environment and the local subsystems leads to final entropies, as:

$$\begin{aligned} S_{\mathcal{G}_E'} &= S_{\mathcal{G}_E} + \Delta S_{E_i^*} + \Delta S_{B_j^*} \\ &= S_{\mathcal{G}_E} + \left(S_{E_i^*} - S_0\right) + \left(S_{B_j^*} - S_0\right) \\ &= S_{\mathcal{G}_E} + S_{E_i^*} + S_{B_j^*} - 2S_0, \end{aligned} \qquad (36)$$

where $\Delta(\cdot)$ stands for the difference, $S_{E_i^*} + S_{B_j^*} > 2S_0$, and the operating entropies of the "post causality-cancellation" subsystems $E_i'$ and $B_j'$ are as follows:

$$\begin{aligned} S_{E_i'} &= S_{E_i} - \Delta S_{E_i^*} \\ &= S_{E_i} - S_{E_i^*} + S_0, \end{aligned} \qquad (37)$$

and

$$\begin{aligned} S_{B_j'} &= S_{B_j} - \Delta S_{B_j^*} \\ &= S_{B_j} - S_{B_j^*} + S_0. \end{aligned} \qquad (38)$$

Without loss of generality, the causality-cancellation of $\mathcal{G}_E$ causes an entropy reduction in the local subsystems $E_i$ and $B_j$ and entropy increment in the $\mathcal{G}_E'$ gravity environment; thus,

$$S_{E_i'} < S_{E_i}, \; S_{B_j'} < S_{B_j} \qquad (39)$$

and

$$S_{\mathcal{G}_E'} > S_{\mathcal{G}_E}. \qquad (40)$$

Because, by theory, the total entropy cannot be decreased in the causality-cancellation, it leads to the following equation:

$$S_{\mathcal{G}_E} + S_{E_i} + S_{B_j} = S_{\mathcal{G}_E'} + S_{E_i'} + S_{B_j'}, \qquad (41)$$

which is satisfied by Equations (36)–(38).

Specifically, we show that the effects of the non-fixed causality on the quantum gravity environment are precisely related to the Landauer principle of the Rindler horizons [11]. This statement is rooted in the fact that at a regulator $d_{\mathcal{G}_E}$, the $D_{\mathcal{G}_E}^{(d_{\mathcal{G}_E})}$ causal development of the quantum gravity environment $\mathcal{G}_E$ is a Rindler horizon $L_{\mathcal{G}_E}$ (see Theorem 1). We step forward from this point.

In particular, a Rindler observer $\Theta$ in the quantum gravity space $\mathcal{G}_E$ sees a vacuum state as follows:

$$\rho_0 = \tfrac{1}{Z} e^{-H_0/k_B T_0}, \qquad (42)$$

where $H_0$ is the corresponding Hamiltonian in the Rindler coordinates, $k_B$ is the Boltzmann constant, $Z$ is the partition function, and $T_0$ is the Unruh temperature; in particular,

$$T_0 = \frac{\hbar a}{2\pi k_B c}, \tag{43}$$

where $a$ is the acceleration parameter [11]. The local subsystems can be modeled as excited states $\rho_{E_i}$ and $\rho_{B_j}$ of a quantum field with entropies $S_{E_i} = -Tr\left(\rho_{E_i} \log_2 \rho_{E_i}\right)$ and $S_{B_j} = -Tr\left(\rho_{B_j} \log_2 \rho_{B_j}\right)$. Specifically, at the causality-cancellation, the quantum gravity environment $\mathcal{G}_E$ absorbs the entropy proportions of

$$S_{E_i^*} = -Tr\left(\rho_{E_i^*} \log_2 \rho_{E_i^*}\right) \tag{44}$$

and

$$S_{B_j^*} = -Tr\left(\rho_{B_j^*} \log_2 \rho_{B_j^*}\right) \tag{45}$$

from $S_{E_i}$ and $S_{B_j}$. To quantify the vanishing entropy pieces of $\rho_{E_i}, \rho_{B_j}$ during the causality-cancellation of $\rho_{\mathcal{G}_E}$, the reduced density matrices $\rho_{E_i^*}$ and $\rho_{B_j^*}$ are used.

Without loss of generality, for an accelerating Rindler observer $\Theta$, the causality-cancellation causes an entanglement entropy difference

$$\Delta S_{E_i^*} = S_{E_i^*} - S_0 = \frac{\Delta E_{E_i^*}}{k_B T_{E_i^*}} \tag{46}$$

and

$$\Delta S_{B_j^*} = S_{B_j^*} - S_0 = \frac{\Delta E_{B_j^*}}{k_B T_{B_j^*}}, \tag{47}$$

where

$$\Delta E_x = Tr\left(\rho_x - \rho_0\right) H_0. \tag{48}$$

In particular, from the second law of thermodynamics follows that the entropy of $\mathcal{G}_E$ increases precisely by $\left(\Delta S_{E_i^*} + \Delta S_{B_j^*}\right)$; thus, during the causality-cancellation, the entropy of the quantum gravity environment increases to $S_{\mathcal{G}_E'}$, specifically as:

$$S_{\mathcal{G}_E'} = S_{\mathcal{G}_E} + \left(\Delta S_{E_i^*} + \Delta S_{B_j^*}\right). \tag{49}$$

The entropy difference $\Delta S_{\mathcal{G}_E} = S_{\mathcal{G}_E'} - S_{\mathcal{G}_E}$ of the quantum gravity environment can be rewritten as follows:

$$\Delta S_{\mathcal{G}_E} = \frac{1}{k_B T_{\mathcal{G}_E}} \Delta E_{\mathcal{G}_E}, \tag{50}$$

where $E_{\mathcal{G}_E}$ is the thermal energy of $\mathcal{G}_E$.

Then the first law of thermodynamics precisely leads to the following conclusion:

$$\Delta S_{\mathcal{G}_E} = \Delta S_{E_i^*} + \Delta S_{B_j^*}. \tag{51}$$

A simple energy conservation argument [11] then results in

$$\Delta E_{\mathcal{G}_E} = E_{E_i^*} + E_{B_j^*}, \tag{52}$$

where $E_{E_i^*}$ and $E_{B_j^*}$ are the thermal energies of the contributing subsystems $E_i^*$ and $B_j^*$ absorbed by the quantum gravity environment $\mathcal{G}_E$. The result in Equation (52) can be rewritten as

$$\begin{aligned}\Delta E_{\mathcal{G}_E} &= E_{E_i^*} + E_{B_j^*} \\ &= k_B T_0 \Delta S_{\mathcal{G}_E}.\end{aligned} \tag{53}$$

Equation (50) can be reevaluated precisely as follows:

$$\Delta S_{\mathcal{G}_E} = \tfrac{1}{k_B T_0}\left(E_{E_i^*} + E_{B_j^*}\right); \tag{54}$$

thus,

$$\tfrac{1}{k_B T_0}\left(E_{E_i^*} + E_{B_j^*}\right) = \Delta S_{E_i^*} + \Delta S_{B_j^*}. \tag{55}$$

Then, without loss of generality, the total entropy change then straightforwardly yields the following:

$$\begin{aligned}\Delta S_{tot} &= \Delta S_{\mathcal{G}_E} + \Delta S_{E_i^*} + \Delta S_{B_j^*} \\ &= 2\left(\Delta S_{E_i^*} + \Delta S_{B_j^*}\right) \\ &= 2\left(\frac{\Delta E_{E_i^*}}{k_B T_{E_i^*}} + \frac{\Delta E_{B_j^*}}{k_B T_{B_j^*}}\right).\end{aligned} \tag{56}$$

The model of the smooth entropy transfer of the causality-cancellation of the quantum gravity environment is summarized in Fig. 4.

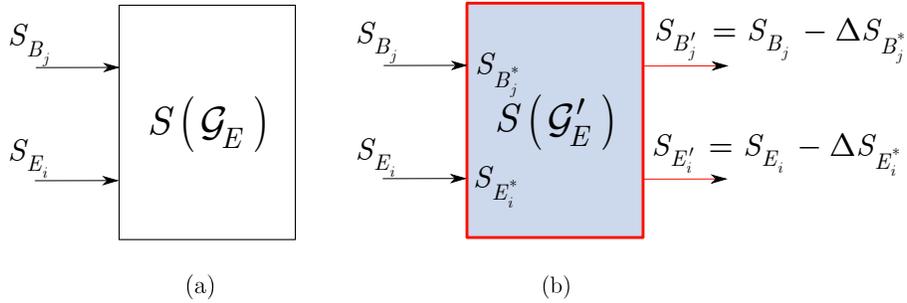

(a)      (b)

**Figure 4.** The smooth entropy transfer of causality-cancellation. (a) The subsystems $B_j, E_i$ of $\rho_{\mathcal{G}_E E_i B_j}$. The density matrices are initialized with entropies $S_{B_j}, S_{E_i}$, and $S_{\mathcal{G}_E}$. (b) The causality-cancellation of $\mathcal{G}_E$ removes the causality relations, and $\mathcal{G}_E'$ has an increased entropy $S_{\mathcal{G}_E'} = S_{\mathcal{G}_E} + S_{E_i^*} + S_{B_j^*} - 2S_0$, whereas the entropies of $B_j', E_i'$ have the reduced entropies of $S_{B_j'} = S_{B_j} - S_{B_j^*} + S_0$, $S_{B_j^*} < S_{B_j}$ and $S_{E_i'} = S_{E_i} - S_{E_i^*} + S_0$, $S_{E_i^*} < S_{E_i}$.

In the second part of the proof, we examine the impacts of the smooth entanglement entropy transfer on a smooth Cauchy slice $\mathcal{C}$. The increased entanglement entropy $S(\mathcal{G}_E')$ of $\mathcal{G}_E'$ leads to an expanded Cauchy area $\mathcal{G}_E'$ on $\mathcal{C}$, which also results in the dilated area of the causal development of $\rho_{\mathcal{G}_E'}$, as follows:

$$D_{\mathcal{G}'_E} = \bigcup_{\forall p_i} \{\mathcal{G}'_E \cap p_i\}. \tag{57}$$

In particular, the restructuration of $\mathcal{C}$ is a direct consequence of the causality-cancellation of $\mathcal{G}_E$. As in the previous case, taking a sufficiently large region $d_{geom}$, $d_{geom} \gg \delta$, the area of $\mathcal{G}'_E$ on $\mathcal{C}$ without loss of generality can be evaluated as follows:

$$\mathcal{G}'^{(d_{geom})}_E = \frac{l^{d-2}}{2\pi} S^{(d_{geom})}_{\mathcal{G}'_E}, \tag{58}$$

where

$$S^{(d_{geom})}_{\mathcal{G}'_E} = \frac{2\pi}{l^{d-2}} \mathcal{G}'^{(d_{geom})}_E. \tag{59}$$

Similar to Theorem 1, it can be exploited that at a regulator $d_{\mathcal{G}_E}$, $\delta \ll d_{\mathcal{G}_E} \ll d_{geom}$, the causal development $D^{(d_{\mathcal{G}_E})}_{\mathcal{G}'_E}$ is a Rindler horizon because within a space-time in a region $\Gamma'$ of size $d_{\mathcal{G}_E}$ near the entangling surface $\Sigma_{\mathcal{G}'_E}$, the space-time looks like a flat space with a Rindler horizon $L_{\mathcal{G}'_E}$; thus,

$$D^{(d_{\mathcal{G}_E})}_{\mathcal{G}'_E} = L_{\mathcal{G}'_E} = \delta D_{\mathcal{G}'_E}, \tag{60}$$

where the areas of the Rindler horizon are changed precisely as follows:

$$\begin{aligned} D^{(d_{\mathcal{G}_E})}_{\mathcal{G}'_E} &> D^{(d_{\mathcal{G}_E})}_{\mathcal{G}_E} \\ &= L_{\mathcal{G}'_E} > L_{\mathcal{G}_E}. \end{aligned} \tag{61}$$

Specifically, at $d_{\mathcal{G}_E} \ll d_{geom}$, by exploiting the constant divergence of the entanglement entropy for $L_{\mathcal{G}'_E}$ and by integrating over the expanded Cauchy area $\mathcal{G}'_E > \mathcal{G}_E$, the leading singularity will also result in the area law.

Thus, the area $\mathcal{G}'_E$ on $\mathcal{C}$ at a non-fixed causality structure can be yielded from the following area law formula [10]:

$$\mathcal{G}'^{(d_{\mathcal{G}_E})}_E = \frac{\delta^{d-2}}{\tilde{c}_0} S^{(d_{\mathcal{G}_E})}_{\mathcal{G}'_E}, \tag{62}$$

where the constant $\tilde{c}_0$ depends on the actual cutoff regulator of the field, and

$$S^{(d_{\mathcal{G}_E})}_{\mathcal{G}'_E} = \frac{\tilde{c}_0}{\delta^{d-2}} \mathcal{G}'^{(d_{\mathcal{G}_E})}_E. \tag{63}$$

The causality-cancellation transforms the Cauchy areas of $E_i$ and $B_j$ onto $E'_i$ and $B'_j$, with local entropies of Equations (37) and (38).

Because for the entropy $S_{\mathcal{G}'_E}$ of the quantum gravity environment $\mathcal{G}'_E$ the relation

$$S_{\mathcal{G}'_E} > S_{\mathcal{G}_E} \tag{64}$$

holds, for the $\Sigma_{\mathcal{G}'_E}$ entangling surface, the relation

$$\Sigma_{\mathcal{G}'_E} > \Sigma_{\mathcal{G}_E} \tag{65}$$

also strictly follows in a non-fixed causality.

At a regulator $d_{\mathcal{G}_E}$, the density $\rho_{\mathcal{G}'_E}$ has the Rindler Hamiltonian; in particular,

$$H_{\mathcal{G}'_E}^{(d_{\mathcal{G}_E})} = 2\pi K_{\mathcal{G}'_E} + c_{\mathcal{G}'_E}$$
$$= -2\pi \int_{x'>0} d^{d-2}y' dx' \left( x' T_{00} \right) + c_{\mathcal{G}'_E}, \quad (66)$$

where $K_{\mathcal{G}'_E}$ is the boost generator in direction $x'$, and $c_{\mathcal{G}'_E}$ is a constant to reach the unit trace in $\rho_{\mathcal{G}'_E}$.

Thus, at $d_{\mathcal{G}_E}$, such that $\delta \ll d_{\mathcal{G}_E} \ll d_{geom}$, the reduced density matrix $\rho_{\mathcal{G}'_E}$ of $\mathcal{G}'_E$ is evaluated as follows:

$$\rho_{\mathcal{G}'_E} = e^{-H_{\mathcal{G}'_E}^{(d_{\mathcal{G}_E})}}. \quad (67)$$

As depicted in Fig. 5, the area of $\mathcal{G}_E$ increases, $\mathcal{G}'_E > \mathcal{G}_E$, along with the area of the causal development at $d_{\mathcal{G}_E}$, $D_{\mathcal{G}'_E}^{(d_{\mathcal{G}_E})} > D_{\mathcal{G}_E}^{(d_{\mathcal{G}_E})}$ (i.e., the area of Rindler horizon of the expanded Cauchy area of the quantum gravity environment).

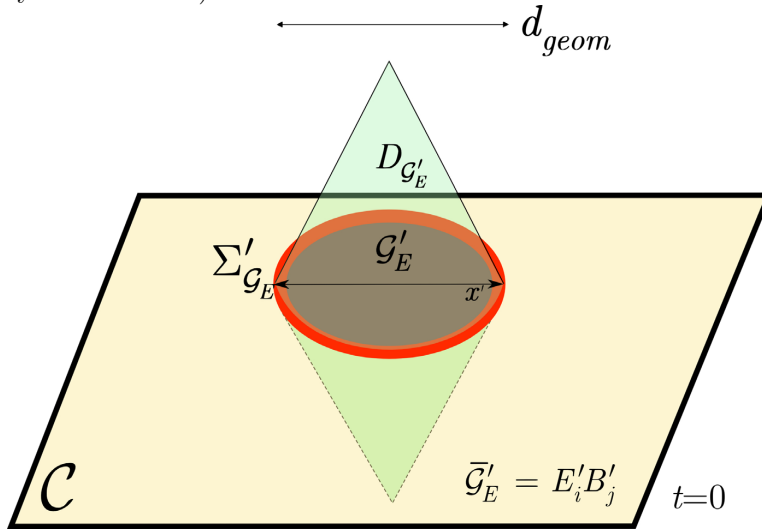

**Figure 5.** The geometric interpretation of the smooth entropy transfer of the causality-cancellation. In a non-fixed causality structure, the Cauchy area of $\mathcal{G}_E$ expands to $\mathcal{G}'_E$, whereas the complementary Cauchy region reduces to $\bar{\mathcal{G}}'_E = E'_i B'_j$. The causal development of $\rho_{\mathcal{G}'_E}$ is $D_{\mathcal{G}'_E}$, with the increased Rindler horizon area $L_{\mathcal{G}'_E} > L_{\mathcal{G}_E}$ at the regulator $d_{\mathcal{G}_E}$, $\delta \ll d_{\mathcal{G}_E} \ll d_{geom}$.

Note that for

$$S_{E^*_i} = S_{E_i},$$
$$S_{B^*_j} = S_{B_j}, \quad (68)$$

the area of $D_{\mathcal{G}'_E}^{(d_{\mathcal{G}_E})}$ is maximized because, in this case, the entropy increase of the quantum gravity environment picks up its theoretical maximum (i.e., $S_{E'_i} = S_{B'_j} = S_0$), resulting in $D_{\mathcal{G}'_E}^{(d_{\mathcal{G}_E})} \to \tilde{L}$, where $\tilde{L}$ is a maximized Rindler horizon.

Some calculations then straightforwardly yield that for
$$c_0 = \tilde{c}_0/4, \tag{69}$$

Equations (62)–(63) can be precisely rewritten via the low-energy field formula of Susskind and Uglum [10] as follows:
$$\begin{aligned}\mathcal{G}'^{(d_{\mathcal{G}_E})}_E &= \tfrac{4}{\Delta\left(\tfrac{1}{G}\right)} S^{(d_{\mathcal{G}_E})}_{\mathcal{G}'_E} \\ &= 4 \tfrac{\delta^{d-2}}{\tilde{c}_0} S^{(d_{\mathcal{G}_E})}_{\mathcal{G}'_E};\end{aligned} \tag{70}$$

thus, for low-energy degrees of freedom of the quantum gravity space, one finds
$$\begin{aligned}S^{(d_{\mathcal{G}_E})}_{\mathcal{G}'_E} &= \tfrac{\mathcal{G}'_E}{4} \Delta\left(\tfrac{1}{G}\right) \\ &= \tfrac{\mathcal{G}'_E}{4} \tfrac{\tilde{c}_0}{\delta^{d-2}}.\end{aligned} \tag{71}$$

Note that Equations (70)–(71) are connected precisely to the Bekenstein-Hawking formula [10]:
$$\Delta S^{(d_{\mathcal{G}_E})}_{\mathcal{G}_E} = \frac{\Delta \mathcal{G}^{(d_{\mathcal{G}_E})}_E}{4G}, \tag{72}$$

where $G$ is the low-energy Newton constant.

∎

**Theorem 3** (Casual development of the quantum gravity environment in a non-fixed causality). *At a regularization $\delta \ll d_{\mathcal{G}_E} \ll d_{geom}$, the vanishing causality structure leads to an increased Rindler horizon area $D^{(d_{\mathcal{G}_E})}_{\mathcal{G}'_E} > D^{(d_{\mathcal{G}_E})}_{\mathcal{G}_E}$, for the quantum gravity environment $\mathcal{G}'_E$.*

*Proof.*
As a first step, we show that the $\Delta S_{\mathcal{G}_E}$ entropy change of the quantum gravity environment causes the Rindler area horizon change. Then we give an exact formula for the change of the Rindler horizon area in a non-fixed causality structure.

Using a regulator $d_{\mathcal{G}_E}$, we can integrate over the Rindler horizon, from which the result of $\Delta E_{\mathcal{G}_E}$, see Equation (53), can be rewritten as follows:
$$\Delta E_{\mathcal{G}_E} = -\kappa \lambda \int_{L_{\mathcal{G}_E}} \mathcal{T}_{ab} \xi^a \Delta L^b_{\mathcal{G}_E}, \tag{73}$$

where $\mathcal{T}_{ab}$ is the matter energy-momentum tensor, and $\xi^a$ is a boost Killing vector field, that is,
$$\xi^a = -\kappa \lambda k_a, \tag{74}$$

where $\kappa$ is the Killing orbit accelerator such that
$$T_0 = \frac{\hbar \kappa}{2\pi k_B c}, \tag{75}$$

and $k_a$ is the tangent vector to the horizon generators with an affine parameter $\lambda$ [11], [55]. In particular, using the formalisms of [11], [55], parameter $\Delta L^b_{\mathcal{G}_E}$ quantifies the area variation of the Rindler horizon $L_{\mathcal{G}_E}$, that is,

$$\Delta L^b_{\mathcal{G}_E} = \xi^b \Delta \lambda \Delta \gamma_{L_{\mathcal{G}_E}}, \tag{76}$$

where $\gamma_{L_{\mathcal{G}_E}}$ is a spatial area element on the cross section of the Rindler horizon $L_{\mathcal{G}_E}$.

Then we rewrite the difference of the expanded $\mathcal{G}'_E$ (see (62)) and the initial $\mathcal{G}_E$ Cauchy areas (see (34)) in terms of Rindler horizon area change at $d_{\mathcal{G}_E}$ as

$$\Delta D^{(d_{\mathcal{G}_E})}_{\mathcal{G}_E} = D^{(d_{\mathcal{G}_E})}_{\mathcal{G}'_E} - D^{(d_{\mathcal{G}_E})}_{\mathcal{G}_E}, \tag{77}$$

which further can be evaluated as

$$\begin{aligned}\Delta D^{(d_{\mathcal{G}_E})}_{\mathcal{G}_E} &= L_{\mathcal{G}'_E} - L_{\mathcal{G}_E} \\ &= -\lambda \int_{L_{\mathcal{G}_E}} \mathcal{R}_{ab} \xi^a \Delta L^b_{\mathcal{G}_E},\end{aligned} \tag{78}$$

where $\mathcal{R}_{ab}$ is the Ricci tensor [55].

Introducing a constant $C$, $\Delta S_{\mathcal{G}_E}$ can be expressed from $\Delta D^{(\delta)}_{\mathcal{G}_E}$, precisely as:

$$\begin{aligned}\Delta S_{\mathcal{G}_E} &= \Delta S_{E_i^*} + \Delta S_{B_j^*} \\ &= C \Delta D^{(\delta)}_{\mathcal{G}_E} \\ &= -\lambda C \int_{L_{\mathcal{G}_E}} \mathcal{R}_{ab} \xi^a \Delta L^b_{\mathcal{G}_E}.\end{aligned} \tag{79}$$

Thus, the expansion of the Rindler horizon area for an accelerating Rindler observer $\Theta$ (i.e., the Rindler horizon area of the quantum gravity environment) is a direct consequence of the causality-cancellation.

The area of the expanded Rindler horizon $L_{\mathcal{G}'_E}$ of $\mathcal{G}'_E$ in a vanishing causality structure, exploiting the derivation in [12], can be expressed precisely as follows:

$$L_{\mathcal{G}'_E} = L_0 + \sqrt{8\pi G} \int_{\mathcal{P}_\infty} d^2 y \int_0^\infty dv v \left(-\psi h_{\mu\nu}\left(x_0\left(\nu\right)\right)\right) k^\mu k^\nu, \tag{80}$$

where $\mathcal{P}_\infty$ is the area of a perturbed surface, $\psi h_{\mu\nu}$ is the wavefunction of the gravitational perturbation, $G$ is Newton's coupling constant, $x_0\left(\nu\right)$ is a trajectory of the perturbed rays, $k$ is the direction parameter of the light rays, and $L_0$ is the area of the section $\mathcal{P}_\infty$ of the beam at $v = 0$. Specifically, using Equation (80), the difference of the Rindler horizons of fixed and non-fixed causality structure quantum gravity environment from Equation (78) can be reevaluated as follows:

$$\begin{aligned}\Delta D^{(\delta)}_{\mathcal{G}_E} &= L_{\mathcal{G}'_E} - L_{\mathcal{G}_E} \\ &= 8\pi G \int d^2 y \int_0^\infty dv v \mathcal{F}_{\mu\nu}(v,y) k^\mu k^\nu,\end{aligned} \quad (81)$$

where $\mathcal{F}_{\mu\nu}$ is the energy-momentum tensor in a flat space [12].

The Rindler horizon area expansion of the quantum gravity space is stimulated by the smooth entanglement entropy transfer, see Equations (36)–(38), leading to the expanded $\mathcal{G}'_E$ Cauchy area and the increased entropy $S_{\mathcal{G}'_E}$ of the quantum gravity environment.

∎

## 5 Conclusions

In this work, we studied the physical fundaments of quantum gravity information processing. Quantum gravity, by integrating the results of general relativity and quantum mechanics, provides us a generalized framework. From the viewpoint of information processing, quantum gravity injects an additional degree of freedom into quantum information processing, by the availability of a non-fixed causality structure. The undetermined causality, in fact, acts as a resource for the communicating parties, which has several consequences both on the information processing structure and on the physics that stands behind information processing. We introduced the term causality-cancellation, which is performed by the quantum gravity space and results in a vanishing causality. We investigated its physical effects and showed that causality-cancellation stimulates the phenomenon of smooth entanglement entropy transfer. We studied the space-time structure of quantum gravity information processing and the impacts of entropy transfer on the space-time geometry. The results indicate that a non-fixed causality structure of quantum gravity information processing has both physical and space-time architectural consequences. This perception can be further exploited in the evolving of quantum gravity computations, particularly in the development of quantum gravity computers and quantum gravity devices.

## Acknowledgements

The results are supported by the grant COST Action MP1006.

# Supplemental Information

## S.1 Notations

The notations of the manuscript are summarized in Table S.1.

**Table S.1.** The summary of the notations.

| Notation | Description |
|---|---|
| $A \rightleftarrows B$ | Parallel realization of the causal directions $A \to B$ and $B \to A$. ($A$ – Alice, $B$ - Bob). |
| $\mathcal{M}_A, \mathcal{M}_B$ | Independent local CPTP maps in the quantum gravity space. |
| $\mathcal{N}_{A_1 B_1}, \mathcal{N}_{A_2 B_2}$ | Local logical channels of $\mathcal{M}_A$ and $\mathcal{M}_B$, defined by Kraus operators $\mathcal{N}_{A_1 B_1}(\rho) = \sum_j A_j^{A_1 B_1} \rho \left( A_j^{A_1 B_1} \right)^\dagger$ and $\mathcal{N}_{A_2 B_2}(\rho) = \sum_j A_j^{A_2 B_2} \rho \left( A_j^{A_2 B_2} \right)^\dagger$. |
| $\mathcal{N}_{A_1 E_1}, \mathcal{N}_{A_2 E_2}$ | Local complementary channels of maps $\mathcal{M}_A$, and $\mathcal{M}_B$, defined by Kraus operators $\mathcal{N}_{A_1 E_1}(\rho) = \sum_j A_j^{A_1 E_1} \rho \left( A_j^{A_1 E_1} \right)^\dagger$ and $\mathcal{N}_{A_2 E_2}(\rho) = \sum_j A_j^{A_2 E_2} \rho \left( A_j^{A_2 E_2} \right)^\dagger$. |
| $B_i, E_i$ | Local output and local environment state of a local CPTP map $\mathcal{M}_i$. |
| $B_j, E_j$ | Remote output and environment state of a remote CPTP map $\mathcal{M}_j$. |
| $\rho_{\mathcal{G}_E E_i B_j}$ | Entangled tripartite qubit system. Defines the entanglement structure of the space-time geometry with local environment $E_i$ and remote output $B_j$. |
| $\rho = \frac{1}{2} \rho_{\mathcal{G}_E E_1 B_2} + \frac{1}{2} \rho_{\mathcal{G}_E E_2 B_1}$ | Density of parallel realizations of local maps $\mathcal{M}_A$, $\mathcal{M}_B$ in a non-fixed causality. |
| $\mathcal{G}_E - E_i B_j,\ E_i - \mathcal{G}_E B_j,\ B_j - \mathcal{G}_E E_i$ | Partitions of $\rho_{\mathcal{G}_E E_i B_j}$. Partition $\mathcal{G}_E - E_i B_j$ is entangled, $E_i - \mathcal{G}_E B_j$, $B_j - \mathcal{G}_E E_i$ are separable. Partition $\mathcal{G}_E - E_i B_j$ models the entangled space-time geometry of the quantum gravity space. |
| $\mathcal{G}_E, E_i, B_j$ | Gravity environment, local environment and remote output, $j \neq i$. Refer also to the corresponding Cauchy regions. |
| $\mathcal{G}'_E, E'_i, B'_j$ | Post causality-cancellation gravity environment, local environment and remote output, $j \neq i$. Refer also to the corresponding Cauchy regions. |
| $\rho$ | Thermal density matrix, $\rho = e^{\frac{-H}{T}}$, where $H$ is the total Hamiltonian, and $T$ is the temperature. |
| $\rho_A$ | Hermitian and semi-definite reduced density matrix, $\rho_A = e^{-H_A}$, where $H_A$ is the modular Hamiltonian. |
| $\rho_0$ | Vacuum state, $\rho_0 = \frac{1}{Z} e^{-H_0/k_B T_0}$ seen by a Rindler observer $\Theta$ within the quantum gravity space $\mathcal{G}_E$, where $H_0$ is the corresponding Hamiltonian in the Rindler coordinates, $k_B$ is the Boltzmann constant, $Z$ is the partition function, while $T_0$ is the Unruh temperature, $T_0 = \frac{\hbar a}{2\pi k_B c}$. |
| $S_A$ | Entanglement entropy between degrees of freedom, $S_A = -Tr(\rho_A \log_2 \rho_A)$. The function is divergent in continuum quantum field theories, thus it requires regulation via a regulator coefficient. |
| $S_A^{(d_{geom})}$ | Entanglement entropy of Cauchy region $A$ at a sufficiently large geometric scale $d_{geom}$ on a Cauchy slice $\mathcal{C}$, $S_A^{(d_{geom})} = \frac{2\pi}{l^{d-2}} A^{(d_{geom})} + \ldots$. |
| $S_A^{(\delta)}$ | Entanglement entropy of Cauchy region $A$ at short distance cut-off regulation $\delta$ on a Cauchy slice $\mathcal{C}$, $S_A^{(\delta)} = c_0 \frac{\tau^{d-2}}{\delta^{d-2}} + c_2 \frac{\tau^{d-4}}{\delta^{d-4}} + \ldots$. |
| $\mathcal{C}$ | Smooth Cauchy slice of a smooth background space-time geometry $\mathcal{B}$. |
| $\mathcal{H}_{tot}$ | Total Hilbert space of the density matrix $\rho_{\mathcal{G}_E E_i B_j}$. The partition $\mathcal{G}_E - E_i B_j$ factorizes the total Hilbert space as $\mathcal{H}_{tot} = \mathcal{H}_{\mathcal{G}_E} \otimes \mathcal{H}_{\bar{\mathcal{G}}_E}$, where $\bar{\mathcal{G}}_E = E_i B_j$. |
| $\Sigma_A$ | Entangling surface of Cauchy region $A$, $\Sigma_A = \partial A$. Entangles the spatial Cauchy region A with the complementary region $\bar{A}$. The area of $\Sigma_A$ is $A$ on a smooth Cauchy slice. |
| $A$ | Cauchy area of the reduced density matrix $\rho_A$. Refers to the degrees of freedom in region $A$ on $\mathcal{C}$. It has a complementary region of $\bar{A}$. |
| $\delta$ | Short-distance cut-off regulation. It regulates the entanglement entropy function in a continuum theory. |
| $d_{geom}$ | A sufficiently large geometric scale in a smooth background space-time of quantum gravity. Interpreted as a geometric length on a smooth Cauchy time slice. |
| $d_{\mathcal{G}_E}$ | A scale regulator for $\mathcal{G}_E$, $\delta \ll d_{\mathcal{G}_E} \ll d_{geom}$. |
| $d$ | Dimension of the space-time. |
| $D_A$ | Causal development of Cauchy region $A$ at $d_{geom}$, $D_A = \bigcup_{\forall p_i} \{A \cap p_i\}$. |
| $l^{d-2}$ | Planck scale in $d$ space-time dimensions, $l^{d-2} = \frac{8\pi G\hbar}{c^3}$, where $G$ is Newton's constant, $\hbar$ is the reduced Planck constant, and $c$ is the speed of light in a vacuum. |
| $\tilde{c}_0$ | A regulator dependent coefficient. |
| $c_k$ | Dimensionless coefficients, sensitive to the regulator of the continuum quantum field theory. |
| $d_A$ | At a regulation $d_A \ll d_{geom}$, a space-time region $\Gamma_A$ of size $d_A$ near the entangling surface $\Sigma_A$ of Cauchy region $A$ can be defined, such that $\delta \ll d_A \ll d_{geom}$. |
| $L_A$ | Rindler horizon of Cauchy region $A$. |
| $\tilde{L}$ | A maximized Rindler horizon, evaluated at $S_{E'_i} = S_{B'_j} = S_0$. |
| $\Gamma_A$ | Space-time region $\Gamma_A$ of size $d_A$. In this region, the space-time looks like a flat space, and the $D_A^{(d_A)}$ causal development of $A$ is a Rindler horizon $L_A$. |
| $D_A^{(d_A)}$ | Causal development of $A$ at regulator $d_A$, $D_A^{(d_A)} = \partial D_A = L_A$. |
| $H_{\mathcal{G}_E}^{(d_{geom})}$ | A sufficiently large scale $d_{geom} \gg \delta$, where $\delta$ is the short-distance cut-off regulator, $H_{\mathcal{G}_E}^{(d_{geom})}$ is a corre- |
| | sponding Hamiltonian. |
| $H_{\mathcal{G}_E}^{(d_{\mathcal{G}_E})}$ | Rindler Hamiltonian at regulator $d_{\mathcal{G}_E}$, where $\delta \ll d_{\mathcal{G}_E} \ll d_{geom}$. |
| $K_{\mathcal{G}_E}$ | Boost generators on the Cauchy slice $\mathcal{C}$. |
| $c_{\mathcal{G}_E}$ | A constant, to reach the unit trace in the reduced density matrix $\rho_{\mathcal{G}_E}$. |
| $\Theta$ | Rindler observer. An inertial Rindler observer is referred as $\Theta_{in}$. |
| $\mathcal{G}_E$, $\bar{\mathcal{G}}_E$ | Cauchy region of the quantum gravity environment, and its complement region, $\bar{\mathcal{G}}_E = E_i B_j$ on a smooth Cauchy slice of a smooth background space-time geometry $\mathcal{B}$. The entangling surface $\Sigma_{\mathcal{G}_E}$ of $\mathcal{G}_E$ generates entanglement between partitions $\mathcal{G}_E$ and $E_i B_j$ of the full density $\rho_{\mathcal{G}_E E_i B_j}$. |
| $\mathcal{G}'_E$ | Expanded Cauchy region of the post causality-cancellation quantum gravity environment $\mathcal{G}'_E$. It has the reduced Cauchy complement region of $\bar{\mathcal{G}}'_E = E'_i B'_j$. |
| $\mathcal{G}_E^{(d_{geom})}$ | Cauchy area of $\mathcal{G}_E$ at scale $d_{geom}$, $\mathcal{G}_E^{(d_{geom})} = \frac{l^{d-2}}{2\pi} S_{\mathcal{G}_E}^{(d_{geom})} + \dots$. |
| $\mathcal{G}_E^{(d_{\mathcal{G}_E})}$ | Cauchy area of $\mathcal{G}_E$ at regulator $d_{\mathcal{G}_E}$, $\mathcal{G}_E^{(d_{\mathcal{G}_E})} = \frac{\delta^{d-2}}{\tilde{c}_0} S_{\mathcal{G}_E}^{(d_{\mathcal{G}_E})}$, where $\delta \ll d_{\mathcal{G}_E} \ll d_{geom}$. |
| $S_{\mathcal{G}_E}^{(d_{geom})}$ | Entanglement entropy of $\mathcal{G}_E$ at scale $d_{geom}$, $S_{\mathcal{G}_E}^{(d_{geom})} = \frac{2\pi}{l^{d-2}} \mathcal{G}_E^{(d_{geom})} + \dots$. |
| $S_{\mathcal{G}_E}^{(d_{\mathcal{G}_E})}$ | Entanglement entropy of $\mathcal{G}_E$ at regulator $d_{\mathcal{G}_E}$, $S_{\mathcal{G}_E}^{(d_{\mathcal{G}_E})} = \frac{\tilde{c}_0}{\delta^{d-2}} \mathcal{G}_E^{(d_{\mathcal{G}_E})}$, where $\delta \ll d_{\mathcal{G}_E} \ll d_{geom}$. |
| $E_i^*, B_j^*$ | These subsystems becomes a part of a thermal system of $\mathcal{G}_E$, during subsystems $E_i, B_j$ are interacting with $\mathcal{G}_E$ in the process of causality-cancellation. |
| $E'_i, B'_j$ | Post causality-cancellation local environment and remote output subsystems. |
| $S_{E'_i}, S_{B'_j}$ | Operating entropies of the local systems $E'_i$ and $B'_j$ in a non-fixed causality, evaluated as |
| | $S_{E'_i} = S_{E_i} - S_{E^*_i} + S_0$, $S_{B'_j} = S_{B_j} - S_{B^*_j} + S_0$. |
| $S_{\mathcal{G}'_E}$ | The entanglement entropy of the quantum gravity environment in a non-fixed causality, $S_{\mathcal{G}'_E} = S_{\mathcal{G}_E} + S_{E^*_i} + S_{B^*_j} - 2S_0$. |
| $\Delta S_{E^*_i}, \Delta S_{B^*_j}$ | Entropy loss of subsystems $E_i, B_j$ in a vanishing causality, $\left(S_{E^*_i} - S_0\right)$, and $\left(S_{B^*_j} - S_0\right)$, where $S_{E^*_i} < S_{E_i}$, $S_{B^*_j} < S_{B_j}$, and $S_0$ is the vacuum entropy. |
| $E$ | Thermal energy change, $\Delta E_x = Tr\left(\rho_x - \rho_0\right) H_0$, where $H_0$ is a corresponding Hamiltonian in the Rindler coordinates. |
| $\Delta E_{\mathcal{G}_E}$ | Thermal energy change of the quantum gravity environment in a vanishing causality, $\Delta E_{\mathcal{G}_E} = E_{E^*_i} + E_{B^*_j}$. |
| $\Delta S_{\mathcal{G}_E}$ | Entanglement entropy change of the quantum gravity environment in a vanishing causality structure, $\Delta S_{\mathcal{G}_E} = \frac{1}{k_B T_{\mathcal{G}_E}} \Delta E_{\mathcal{G}_E}$. |
| $\mathcal{T}_{ab}$ | Matter energy-momentum tensor. |
| $\xi^a$ | A boost Killing vector field, $\xi^a = -\kappa \lambda k_a$. |
| $\kappa$ | The Killing orbit accelerator. |
| $k_a$ | Tangent vector to the horizon generators with an affine parameter $\lambda$. |
| $\Delta L^b_{\mathcal{G}_E}$ | Quantifies the area change of the Rindler horizon, $\Delta L^b_{\mathcal{G}_E} = \xi^b \Delta \lambda \Delta \gamma_{L_{\mathcal{G}'_E}}$, where $\gamma_{L_{\mathcal{G}'_E}}$ is a spatial area element on the cross-section of the Rindler horizon $L_{\mathcal{G}_E}$. |
| $\Delta D^{(d_{\mathcal{G}_E})}_{\mathcal{G}_E}$ | Change of the Rindler horizon area at regulator $d_{\mathcal{G}_E}$, $\Delta D^{(d_{\mathcal{G}_E})}_{\mathcal{G}_E} = D^{(d_{\mathcal{G}_E})}_{\mathcal{G}'_E} - D^{(d_{\mathcal{G}_E})}_{\mathcal{G}_E}$. |
| $\mathcal{R}_{ab}$ | Ricci tensor. |
| $\mathcal{P}_\infty$ | Area of a perturbed surface. |
| $\psi h_{\mu\nu}$ | Wavefunction of the gravitational perturbation. |
| $G$ | Newton's coupling constant. |
| $x_0(\nu)$ | Trajectory of the perturbed rays. |

| $\mathcal{F}_{\mu\nu}$ | Energy-momentum tensor in a flat space. |

## S.2 Abbreviations

| | |
|---|---|
| **CPTP** | **Completely Positive Trace Preserving** |
| **QFT** | **Quantum Field Theory** |
| **UV** | **Ultra Violet** |